\documentclass[twocolumn,prl,amsmath,amssymb,superscriptaddress]{revtex4-1}

\usepackage{amsmath,amssymb,latexsym}
\usepackage{graphicx}
\usepackage{bm}
\usepackage{mathrsfs}
\usepackage{physics}
\usepackage{siunitx}
\usepackage[version=4]{mhchem}
\usepackage[normalem]{ulem}
\usepackage{color}
\usepackage{nicefrac}
\usepackage{hyperref}
\usepackage{array}
\usepackage[mathscr]{euscript}
\usepackage{mathtools}
\usepackage{cases}

\makeatletter
\def\clearfmfn{\let\@FMN@list\@empty}
\makeatother

\begin{document}

\title{Far-from-equilibrium dynamics of angular momentum\\[5pt] in a quantum many-particle system}

\author{Igor N. Cherepanov}
\thanks{These two authors contributed equally}
\affiliation{Institute of Science and Technology Austria, Am Campus 1, 3400 Klosterneuburg, Austria}

\author{Giacomo Bighin}
\thanks{These two authors contributed equally}
\affiliation{Institute of Science and Technology Austria, Am Campus 1, 3400 Klosterneuburg, Austria}

\author{Lars Christiansen}
\affiliation{Department of Chemistry, Aarhus University, 8000 Aarhus C, Denmark}

\author{Anders Vestergaard J{\o}rgensen}
\affiliation{Department of Chemistry, Aarhus University, 8000 Aarhus C, Denmark}

\author{Richard Schmidt}
\affiliation{Max Planck Institute for Quantum Optics, Hans-Kopfermann-Str. 1, 85748 Garching, Germany}
\affiliation{Munich Center for Quantum Science and Technology, Schellingstra{\ss}e 4, 80799 M\"unchen, Germany}

\author{Henrik Stapelfeldt}
\email[Corresponding author: ]{henriks@chem.au.dk}
\affiliation{Department of Chemistry, Aarhus University, 8000 Aarhus C, Denmark}

\author{Mikhail Lemeshko}
\email[Corresponding author: ]{mikhail.lemeshko@ist.ac.at}
\affiliation{Institute of Science and Technology Austria, Am Campus 1, 3400 Klosterneuburg, Austria}

\date{\today}

\begin{abstract}

We use  laser-induced rotation of single  molecules embedded in superfluid helium nanodroplets to reveal angular momentum dynamics and transfer in a controlled setting, under far-from-equilibrium conditions. As an unexpected result, we observe pronounced oscillations of time-dependent molecular alignment that have no counterpart in gas-phase molecules. Angulon theory reveals that these oscillations originate from the unique rotational structure of molecules in He droplets and quantum-state-specific transfer of rotational angular momentum to the many-body He environment on picosecond timescales. Our results pave the way to understanding collective effects of macroscopic angular momentum exchange in solid state systems in a bottom-up fashion.

\end{abstract}

\maketitle


Revealing microscopic dynamics of angular momentum in solids is of key importance for designing molecular magnets~\cite{CaleroPRL05}, spintronic  and nano-magneto-mechanic devices~\cite{KovalevPRB2007,MatsuoPRL11, KovalevPRL05,Tejada2010PRL}, ultrafast magnetic switches and data registers~\cite{BeaurepairePRL96,KoopmansPRL2005,KirilyukRMP10}, as well as for controlling decoherence in  solid-state qubits~\cite{BassettScience14,DonatiScience16}. Experimentally, being able to fine-tune the relative strength of the spin--orbit, electron--electron, and electron--lattice couplings,  would allow one to separate their relative contributions to angular momentum dynamics in magnetic systems. However, achieving such degree of control is beyond the reach of most solid-state experiments. Theoretically, due to the non-Abelian algebra describing quantum rotations, the problem of angular momentum dynamics becomes seemingly intractable for systems of many particles~\cite{VarshalovichAngMom}.

Here we use a  controllable quantum many-body system -- isolated molecules trapped in nanodroplets of superfluid helium --   to study out-of-equilibrium angular momentum dynamics. In experiment, we use a  picosecond laser pulse  to suddenly align the molecules   and thereby bring the system  far away from equilibrium.  As a  novel and unexpected result, we observe pronounced oscillations  in  time-dependent molecular alignment,  measured with a femtosecond probe pulse, that have no counterpart in gas-phase molecules. Theoretically, we develop a finite-temperature quantum many-body theory based on angulon quasiparticles, which explains the microscopic origins of this phenomenon.

Experimentally, 10-nm-diameter helium droplets, each doped with at most one  iodine (\ce{I_2}),  carbon disulfide (\ce{CS_2}) or carbonyl sulfide (\ce{OCS}) molecule, are first irradiated by a \SI{15}{ps} linearly polarized alignment laser pulse, which excites a  superposition of molecular rotational states and thus provides the molecules with a tunable amount of rotational angular momentum. This allows us to explore the energy and lifetime of highly excited angular momentum states, inaccessible through conventional infrared and microwave spectroscopies, typically restricted to the linear response regime. After a time delay, $t$, the spatial orientation of the molecules is measured by Coulomb explosion with a \SI{40}{fs} intense probe pulse and recording of the emission direction of fragment ions. Hereby, $\langle \cos^2  \theta_\text{2D} \rangle$,  the standard measure for the degree of the molecular alignment is determined, $\theta_\text{2D}$  being the angle between the alignment pulse polarization and the projection of the velocity vector of a fragment ion on the detector~\cite{sup}.


\begin{figure}[t]
\centering
\includegraphics[width=1.\linewidth]{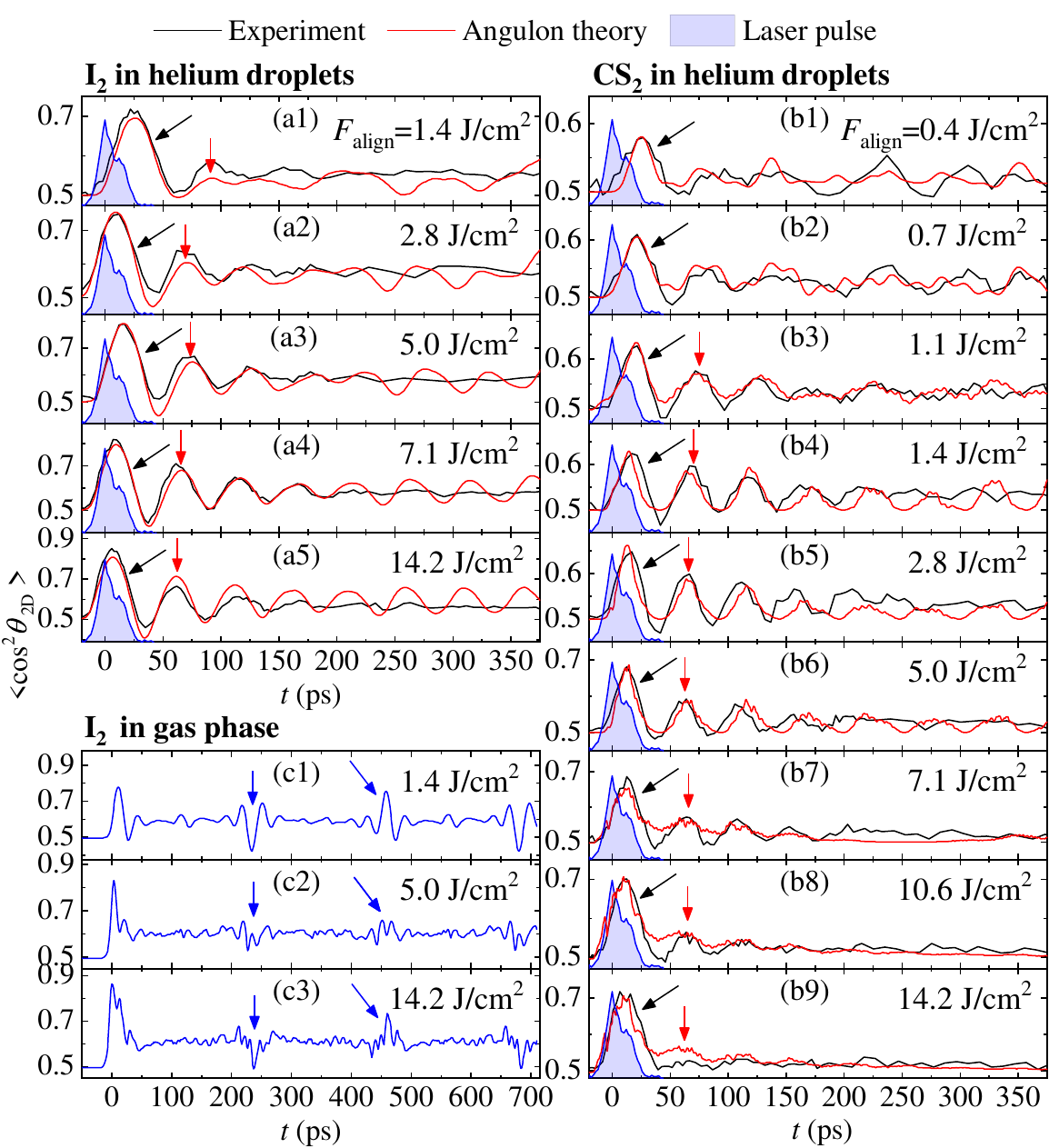}
\caption{\label{fig:cos} (a), (b)~Time evolution of the  degree of alignment, $\langle \cos^2  \theta_\text{2D} \rangle$, for \ce{I2} and \ce{CS_2}  in He droplets induced by a \SI{15}{ps} laser pulse (intensity profile shown in blue)  with fluence $F_\text{align}$.   Experiment (black lines) is compared to   finite-temperature angulon theory (red  lines). (c)~Calculated time evolution of $\langle \cos^2 \theta_\text{2D} \rangle$ for gas-phase \ce{I2} molecules. }
\end{figure}

Figure~\ref{fig:cos} shows $\langle \cos^2 \theta_\text{2D} \rangle$ as a function of time for  \ce{I2} and \ce{CS_2} and various fluences of the alignment pulse, $F_\text{align}$. For \ce{I_2}, Fig.~\ref{fig:cos}(a), a distinct peak in $\langle \cos^2 \theta_\text{2D} \rangle$ is observed shortly after the alignment pulse at all fluences. This peak (marked by black arrows) grows in magnitude and shifts to earlier times as  $F_\text{align}$ is increased,  consistent with previous observations of laser-induced alignment of molecules in He droplets~\cite{PentlehnerPRL13,Shepperson16}.    The striking new observation is,  however, that the distinct oscillations in $\langle \cos^2 \theta_\text{2D} \rangle$ with a period of $\sim \SI{51}-\SI{56}{ps}$ (red arrows in Fig.~\ref{fig:cos}) that follows the prompt alignment peak. The oscillations are already visible at $F_\text{align} = \SI{1.4}{J/cm^2}$ and get substantially more pronounced at higher fluences, reaching a maximum for $F_\text{align} = \SI{7.1}{J/cm^2}$. At $F_\text{align} = \SI{14.2}{J/cm^2}$ the oscillation magnitude slightly  decreases  again.

A similar effect was observed for \ce{CS_2} molecules, Fig.~\ref{fig:cos}(b). Here, the measurements were extended to a broader range of fluences, showing the emergence of oscillations at around $F_\text{align} = \SI{0.7}{J/cm^2}$, their growth in magnitude with increasing fluence reaching a maximum around $F_\text{align} = \SI{2.8}{J/cm^2}$, and their gradual disappearance for $F_\text{align} \geq \SI{7.1}{J/cm^2}$. Strikingly, the period of the oscillations ($\sim \SI{46}{ps}$) is essentially constant for the $\langle \cos^2 \theta_\text{2D} \rangle$ traces recorded with $F_\text{align}$ between 2.8 and \SI{10.6}{J/cm^2}. Finally, similar measurements  on \ce{OCS} reveal much less pronounced oscillations compared to \ce{I_2} and \ce{CS_2}~\cite{sup}.

For comparison, Fig.~\ref{fig:cos}(c) shows the calculated alignment dynamics for gas-phase \ce{I2} molecules, revealing the characteristic half-and-full revivals (marked by blue arrows) of the rotational wave packet. The period and magnitude of the pronounced oscillations of Fig.~\ref{fig:cos}(a) are completely different from that of the gas-phase revivals. This hints at a novel non-perturbative mechanism, which cannot be  understood qualitatively through previously known gas-phase phenomena.

Next, we use the angulon quasiparticle theory~\cite{SchmidtLem15,LemeshkoPRL17} to explain  the experimental observations. A slowly rotating linear molecule interacting with a bosonic bath can be described  by the Hamiltonian~\cite{SchmidtLemPRX16}:
\begin{equation}
\hat{H}=B(\boldsymbol{\mathrm{\hat{L}}}-\boldsymbol{\mathrm{\hat{\Lambda}}})^2 +\sum_{k \lambda \mu} \omega_k \hat{b}^{\dagger}_{k \lambda \mu}\hat{b}_{k \lambda \mu}+\sum_{k \lambda} V_{k \lambda}\big(\hat{b}^{\dagger}_{k \lambda 0} +\hat{b}_{k \lambda 0}\big) \; .
\label{eq:h}
\end{equation}
Here $B = \hbar^2/(2 I)$ is the molecular rotational constant, with $I$ the molecular moment of inertia, and $\boldsymbol{\mathrm{\hat{L}}}$  the total angular-momentum operator of the combined system, consisting of a molecule and  helium excitations. $\boldsymbol{\mathrm{\hat{\Lambda}}}$  is the angular-momentum operator for the bosonic helium bath, whose excitations are described by the creation (annihilation) operators, $\hat{b}^{\dagger}_{k \lambda \mu}$ ($\hat{b}_{k \lambda \mu}$), respectively. For convenience, the creation and annihilation operators are cast in the angular momentum basis~\cite{LemSchmidtChapter}, with $k$ the magnitude of linear momentum of the boson, $\lambda$ its angular momentum, and $\mu$ the angular momentum projection onto the $z$-axis. $\omega_k$ in Eq.~\eqref{eq:h} gives the dispersion relation of superfluid helium and $V_{k \lambda}$ encodes the details of the molecule-helium interactions~\cite{sup}.

The interaction with a linearly polarized far-off-resonant laser  is modelled as $\hat{H}_\text{laser} = - \frac{1}{4} \Delta \alpha E^2 (t) \cos^2 \hat{\theta}$,
where $\Delta \alpha$ is the molecular polarizability anisotropy, $E(t)$ is the laser electric field  polarized in the $z$-direction, and $\hat{\theta}$ is the operator describing the angle between the molecular axis and the laboratory $z$-axis~\cite{LemKreDoyKais13}.  We describe the many-body dynamics in the absence of the laser  using the time-dependent variational ansatz \cite{Kramer:1981} characterized by the quantum numbers $L$, the total angular momentum, and $M$, the projection onto the laboratory-frame $z$ axis:
\begin{multline}
\ket{\psi_{LM, i} (t)} =  \hat{U} \biggl ( g_{LM}(t) \ket{LM0}_\text{mol} +  \\ + \sum_{k \lambda n} \alpha_{k \lambda n}(t) \ket{LMn}_\text{mol} \hat{b}^{\dagger}_{k \lambda n} \biggr)  \ket{i}_\text{bos}
\label{eq:tdva}
\end{multline}
Here $\ket{i}_\text{bos}$  represents  the many-body states of the helium bath, and $g_{LM}$ and $ \alpha_{k \lambda n}$ are time-dependent variational parameters.  $\hat{U} = \exp \left[ \sum_{k \lambda} V_\lambda(k) /(\omega_k + B \lambda (\lambda + 1)) (\hat{b}_{k \lambda 0} - \text{h.c.}) \right ]$
is a coherent-state transformation that diagonalises the Hamiltonian (\ref{eq:h})  in the limit of a slowly-rotating impurity, $B \to 0$. Such a transformation excites an infinite number of bosons, accounting for a macroscopic deformation of the bath by the molecular impurity \cite{SchmidtLemPRX16}, on top of which Eq.~\eqref{eq:tdva} takes into account additional single-phonon excitations.  As a laser creates superpositions of states with different $L$, an appropriate time-dependent variational ansatz to describe the evolution of the system as described by the full Hamiltonian, $\hat{H}+\hat{H}_\text{laser}$, is $\ket{\Psi_i (t)} = \sum_{LM} \ket{\psi_{LM, i} (t)}$.

The finite-temperature Lagrangian is defined as a thermal expectation value, $\mathscr{L}(t) = Z_\text{bos}^{-1} \sum_i e^{-\beta E_i} \matrixel{\Psi_i (t)}{\mathrm{i} \partial_t - \hat{H}  - \hat{H}_\text{laser}}{\Psi_i (t)}$, where $E_i$ is the energy of the $\ket{i}$ bath state, and $Z_\text{bos} \equiv \sum_i \exp (-\beta E_i)$ is  the partition function accounting for the finite temperature of the bath~\cite{DeAngelis:1991dj}. The finite temperature of the molecule, on the other hand, is included through the averaging over the statistical mixture of initial equilibrium configurations, each state being weighted by the spin statistics and Boltzmann factors~\cite{sup}. Time evolution of the states $\vert \Psi_i  (t)\rangle$ is obtained by numerically solving the Euler-Lagrange equations corresponding to $\mathscr{L} (t)$. In addition, in order to account for the effect of centrifugal distortion relevant for high angular momentum states~\cite{Lehmann:2001gr,Harms:1997iv,Hartmann:1995cr}, we introduce a phenomenological term, $ - D [L(L+1)]^2$, in the equations of motion, with $D$ the centrifugal distortion constant~\cite{sup}.

The degree of alignment calculated using the angulon theory, shown in Fig.~\ref{fig:cos}(a)-(b) by the red curves, reproduces the main features observed in the experiment. At early  times, the theory describes the prompt alignment peak for all three molecules.  Most important, the theory reproduces the oscillations observed in the experiment for I$_2$ and CS$_2$. In line with the experimental findings, the magnitude of these oscillations gradually increases with fluence, and then starts decreasing. Persistent oscillations present for I$_2$ at $F_\text{align} = 7.1$ and $\SI{14.2}{J/cm^2}$ at long times in the theoretical curves and absent in the  experiment, most likely arise due to the Hilbert space truncation enforced by our ansatz~\cite{Parish:2016gm}.

\begin{figure}[t]
\centering
\includegraphics[width=0.9\linewidth]{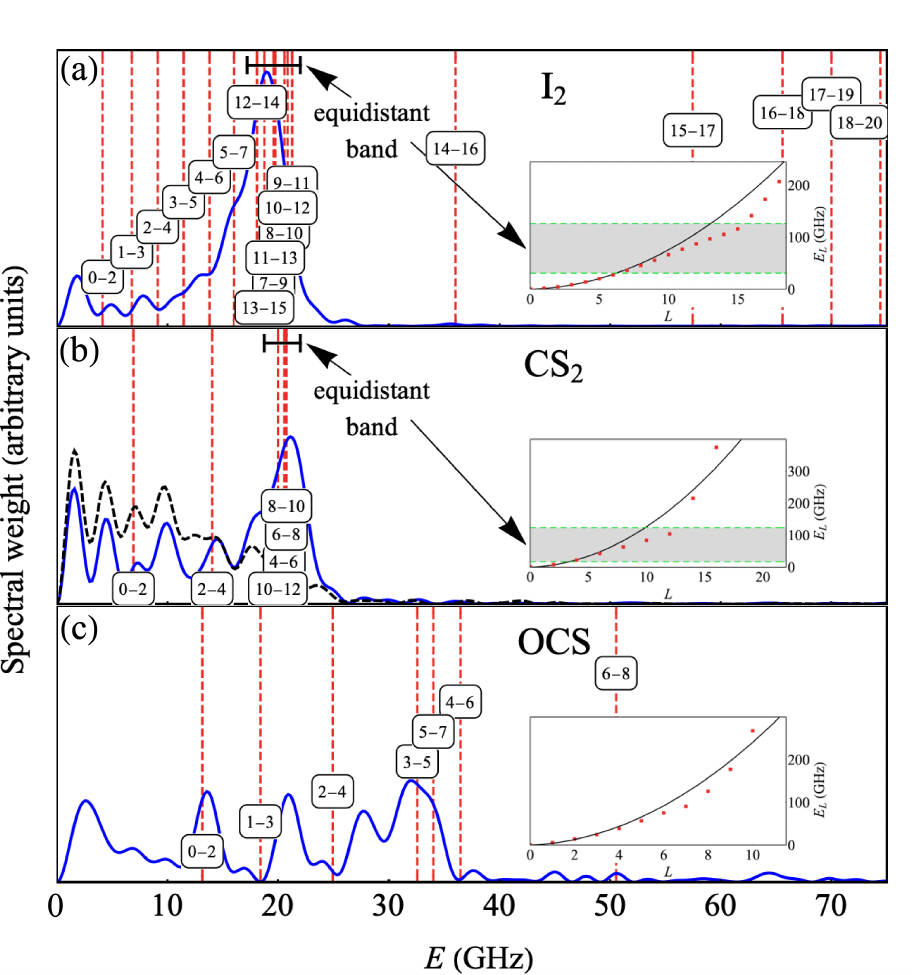}
\caption{\label{fig:spectrum} (a)~Fourier transform of the $\langle \cos^2 \theta_\text{2D} \rangle$ trace  for \ce{I_2} at $F_\text{align}=\SI{7.1} {J/cm^2}$ (blue). Vertical red dashed lines denote the theoretical wave packet frequencies, $\nu_{L,L'}$. Inset: theoretically derived  rotational energy levels  (red dots) compared with a $B^* L(L+1)$ interpolation  (black  line), making the role of centrifugal distortion apparent. (b)~Same as (a), but for \ce{CS_2} at $F_\text{align}=\SI{7.1} {J/cm^2}$ (blue) and  $\SI{14.2} {J/cm^2}$ (black dashes).  (c)~Same as (a) but for \ce{OCS} at $F_\text{align}=\SI{2.8} {J/cm^2}$.}
\end{figure}

However, what is the origin of the oscillations?   As detailed below, the oscillations  are the result of two conditions caused by the superfluid helium environment: (i)~the rotational spectrum of the molecules differs strongly from that of the gas phase case, and (ii)~rotational angular momentum of the molecules is transferred to the elementary excitations in the He droplet on picosecond timescales.

To get insight into the rotational structure of the molecules we analyze the  Fourier transform of the experimentally measured  $\langle \cos^2 \theta_\text{2D} \rangle$ traces.  The  blue solid lines in Figs.~\ref{fig:spectrum}(a) and (b) shows the results for \ce{I_2} and \ce{CS2}, at $F_\text{align} = \SI{7.1}{J/cm^2}$. In both cases the spectrum is  dominated by a single peak,  centered at $\sim \SI{19}{GHz}$ for \ce{I_2}  and at $\sim \SI{22}{GHz}$ for \ce{CS_2}. These frequencies naturally correspond to the oscillation periods of  $\sim \SI{51}{ps}$ in Fig.~\ref{fig:cos}(a4) and of $\sim \SI{46}{ps}$ in Fig.~\ref{fig:cos}(b7). In contrast, for \ce{OCS} the power spectrum does not contain a single dominant peak. As a result, the oscillations observed in $\langle \cos^2 \theta_\text{2D} \rangle$~\cite{sup} are less pronounced compared to those for \ce{I2} and \ce{CS2}.

The molecule-laser interaction creates a rotational wave packet, i.e.\ a coherent superposition  of  states (each composed of a molecular rotational state dressed by phonons), whose  total angular momentum differs by $L-L' = \pm 2, 0$ and with $M = M'$~\cite{LemKreDoyKais13}. Peaks in the power spectrum of Fig.~\ref{fig:spectrum} reflect the coherences between such states.  The vertical dashed lines in Fig.~\ref{fig:spectrum} mark the frequencies, $\nu_{L,L'} = (E_L - E_{L'})/h$, corresponding to these coherences, where the rotational energies $E_L$ aew obtained from the angulon theory. For \ce{I_2} (\ce{CS2}) seven (four) frequencies cluster around $\SI{18}-\SI{20}{GHz} (\SI{21}{GHz})$ and thereby explains the origin of the dominant peak in the power spectra and the pronounced $\langle \cos^2 \theta_\text{2D} \rangle$ oscillations. For \ce{OCS} the 3-5 4-6 and 5-7 frequencies only lie fairly close and, consequently, the $\langle \cos^2 \theta_\text{2D} \rangle$ oscillations become less pronounced~\cite{sup}.

The reason for the clustering of the frequencies is the effect of the large centrifugal distortion constant of molecules in He droplets~\cite{ChoiIRPC06,LehnigFarDiss09}. To illustrate this explicitly, the insets in Fig.~\ref{fig:spectrum} compare the calculated rotational energies for molecules in He droplets with the centrifugal distortion constant included (red dots) and not included (black lines). The energies from the calculation including centrifugal distortion differs strongly from the pure rigid rotor energies. This stands in stark contrast to the case of gas phase molecules where the centrifugal term is only a small perturbation to the rigid rotor structure except for superrotor states accessed by optical centrifuges~\cite{milner_probing_2017}.

We have demonstrated that the pronounced $\langle \cos^2 \theta_\text{2D} \rangle$ oscillations for \ce{I2} and \ce{CS2} (Fig.~\ref{fig:cos}) results from the presence of a band of equidistant states (insets in Fig.~\ref{fig:spectrum}). However, why are these states populated in such a robust manner?  Notably, why are the oscillations almost identical in the broad range of intermediate laser fluences, Fig.~\ref{fig:cos}(b3-b7), but disappear for  the weakest and strongest pulses, Fig.~\ref{fig:cos}(b1) and (b9)? 

First of all, a particular feature of  the  rotational spectra shown in Figs.~\ref{fig:spectrum}(a)-(b) is a large energy gap in the frequencies of the coherences right after the dominant peaks at $\sim \SI{19}{GHz}$ and $\sim \SI{22}{GHz}$ (For CS2 the next frequency after the $\sim \SI{22}{GHz}$ peak is $\nu_{12,14} = \SI{111}{GHz}$ outside the range of Fig.~\ref{fig:spectrum}(b)). This gap plays an important role in stabilizing the pronounced oscillations for a broad range of intermediate laser fluences,  ranging from $F_\text{align} = 2.8$ to $\SI{14.2}{J/cm^2}$ for \ce{I2} (Fig.~\ref{fig:cos}(a2--a5)), and from $F_\text{align} = 1.1$ to $\SI{7.1}{J/cm^2}$ for \ce{CS2} (Fig.~\ref{fig:cos}(b3--b7)). In the case of a Gaussian alignment pulse, treated as a perturbation $\hat{V}(t)$ to first order, the transition probability between discrete stationary states, $L$ and $L'$, is given by  $W_{L,L'}=\mathrm{exp} (-\sigma^2 \nu_{L,L'}^2) |V_{L,L'}|^2 / \hbar^2$, where $\sigma$ is the pulse duration~\cite{sup}. The exponential factor in $W_{L,L'}$, which, in the spirit of Fermi's golden rule, we interpret as a discrete analogue of the phase space density, strongly suppresses any transfer of spectral weight beyond the band of equidistant states, i.e. the gap effectively creates a barrier in angular momentum space. In fact, when $F_\text{align}$ is increased to $\SI{14.2}{J/cm^2}$, the spectral weight is even transferred to lower angular momentum states, as shown in Figs.~\ref{fig:spectrum}(b) by the black dashed line. This effect is reminiscent of bouncing a wave packet against a wall, which, here, promotes population of states with lower energies and thereby lead to disappearance of the oscillations - see \ref{fig:cos}(b9). Such a behavior is qualitatively different from the case of  gas-phase molecules, where increasing the laser intensity always promotes the population towards higher angular-momentum states~\cite{seideman_dynamics_2001}.

\begin{figure}[t]
\centering
\hspace{-34pt} \includegraphics[width=0.9\linewidth]{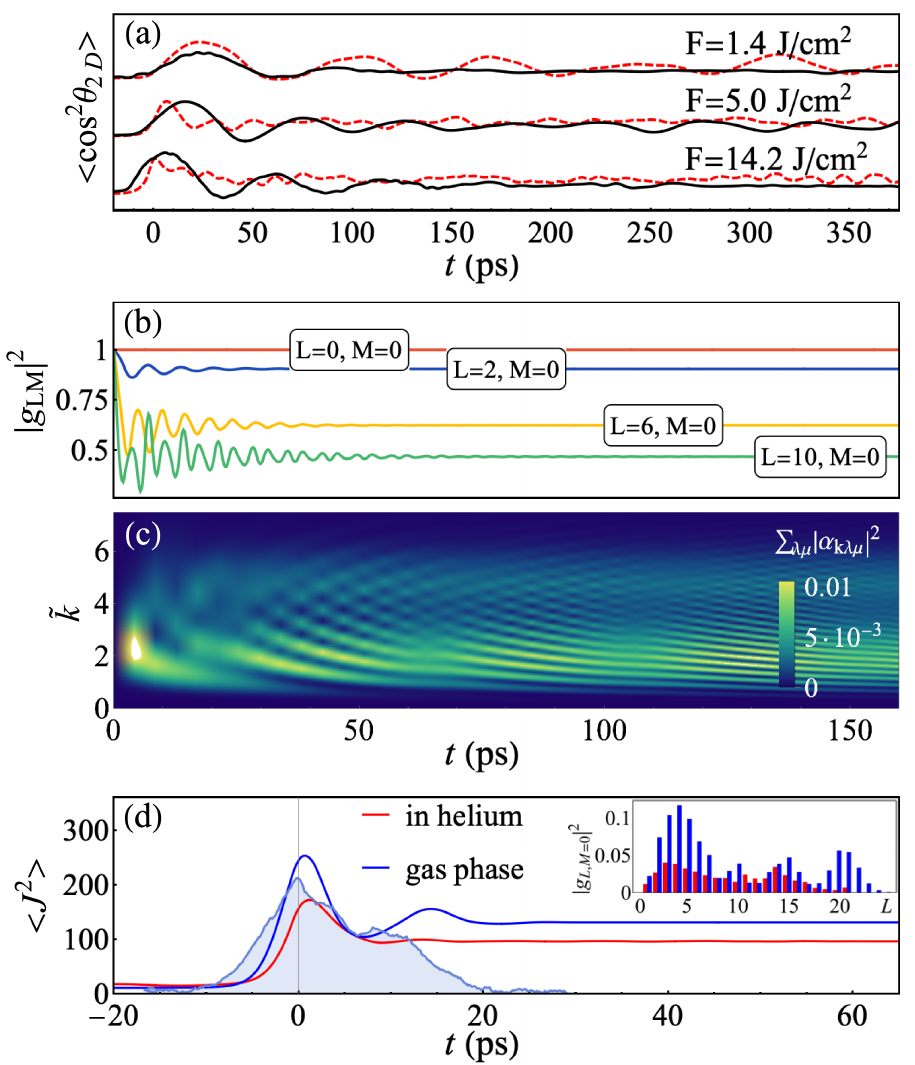}
\caption{\label{fig:fig3} (a)~Theoretical time evolution of $\langle \cos^2 \theta_\text{2D} \rangle$ for \ce{I2} neglecting the dynamical transfer of angular momentum (red dashed line) is not able to describe the experimental observations (black solid line). (b)~Time evolution of the molecular  populations, $|g_{LM}|^2$, for $L=0,2,6,10$ and $M=0$.  (c)~Total phonon population, $\sum_{\lambda \mu} |\alpha_{k \lambda \mu}|^2$ (for the state with $L=2$), as a function of time and of the (dimensionless) momentum, $\tilde{k} = k (m_\text{He}B)^{-\nicefrac{1}{2}}$, $m_\text{He}$ being the mass of a helium atom. (d)~Time evolution of the molecular angular momentum, $\langle \boldsymbol{\mathrm{\hat{J}}}^2 \rangle \equiv \langle(\boldsymbol{\mathrm{\hat{L}}}-\boldsymbol{\mathrm{\hat{\Lambda}}})^2 \rangle$ in helium (red) and in the gas phase (blue). Inset: molecular populations $|g_{LM=0}|^2$ as a function of $L$, in helium (red) and in the gas phase (blue), at $t = 160$~ps.}
\end{figure}

The rotational energy structure of molecules in superfluid helium is, however, not the only effect responsible for the oscillation dynamics in $\langle \cos^2 \theta_\text{2D} \rangle$. Indeed, Fig.~\ref{fig:fig3}(a) shows the alignment simulations for \ce{I_2} when the dynamical transfer of angular momentum between the molecule and the many-body helium bath is neglected. (red dashed line). Poor agreement with experimental results (black line) suggests such transfer plays a crucial role.

Let us start by answering the following question: assuming that we instantaneously switch on the molecule-helium interactions, how long does it take for a molecule to equilibrate with the helium environment and form an angulon quasiparticle? In Fig.~\ref{fig:fig3}(b), we present the time evolution of molecular rotational state populations,  $|g_{LM}|^2$, cf.\ Eq.~\eqref{eq:tdva}, for  \ce{I_2} after its instantaneous immersion in superfluid helium (in the absence of a laser field).  Crucially, we observe that the equilibrium value of  $|g_{LM}|^2$, reached at long times, decreases as the initial angular momentum $L$ increases.  Similarly, the equilibration time is also $L$-dependent. In other words, the angulon quasiparticle weight, $|g_{LM}|^2$, decays during the first picoseconds of evolution, and is transferred to the population of phonon amplitudes, $\alpha_{k \lambda n}$ of Eq.~\eqref{eq:tdva}, resulting in a superposition of angular momentum of the molecule and an excitation in helium, see Fig.~\ref{fig:fig3}(c). Such an equilibration time scale, reflecting the time scales of the molecule--He interactions, is always  on the order of tens of ps, which is  comparable to the laser pulse duration. This fact has far-reaching consequences: when the  alignment pulse  is  on, it pushes a fraction of the  rotational wave packet to high $L$-states, thereby increasing the respective molecular populations in $|g_{LM}|^2$. At the same time, the molecule-bath interaction counteracts by creating a field of excitations in the superfluid helium around the molecule, thereby decreasing  $|g_{LM}|^2$. Since these two competing processes happen on the same timescale their interplay in the first few tens of ps is crucial in determining the long-time alignment dynamics.

The delicate interplay between laser- and bath-induced dynamics is confirmed by Fig.~\ref{fig:fig3}(d), which shows the  time evolution of $\langle \boldsymbol{\mathrm{\hat{J}}}^2 \rangle$, where $ \boldsymbol{\mathrm{\hat{J}}} = \boldsymbol{\mathrm{\hat{L}}}-\boldsymbol{\mathrm{\hat{\Lambda}}}$ is the rotational angular momentum of the molecule  alone (disregarding the angular momentum of the phonon cloud), see Eq.~\eqref{eq:h}, for I$_2$ at $F=7.1$ J/cm$^2$.  Already  after just a few ps, when mostly states with low $L$'s are populated, the effect of superfluid helium on the angular momentum dynamics becomes apparent, as it prevents  the rotational angular momentum of the molecule from increasing  as rapidly as it would in the gas phase. The inset of Fig.~\ref{fig:fig3}(d), shows   $|g_{LM}|^2$  as a function of $L$  after the pulse for the gas phase and in helium. One can see that for \ce{I_2} in He droplets  the $L>15$ states are essentially not populated. This  further supports the `barrier' picture introduced above.

Finally, we mention that if the  alignment pulse acted for a period considerably shorter than the molecule-helium interaction timescale, the level structure would not be modified by the dynamical many-body dressing, even at large fluences, thereby making the state clustering of Fig.~\ref{fig:spectrum} much less pronounced.  In that case the bath-induced dynamics would not have the time to follow the faster laser-induced dynamics, so that  non-equilibrated states beyond the  band of equidistant states could be populated, resulting in the  absence of oscillations. This situation has been observed \cite{Shepperson16}, and can be interpreted as `detachment'  of the molecule from the surrounding helium shell.

Similarly to how the understanding of phonon-mediated superconductivity developed from a description of the dressing of single electrons  by phonons in solids~\cite{EsterlisNQM18}, our results can pave the way, within a bottom-up-approach, to understanding the collective effects of macroscopic angular momentum exchange in condensed matter systems. In contrast to the past studies of angular momentum transfer in collisions between molecular beams~\cite{OnvleeNatChem17,KleinNatPhys17}, our results shed light on the influence of a solvent as well as  on the timescales of angular momentum transfer.  Finally, in the spirit of quantum simulation in ultracold  gases~\cite{GrossScience17}, paramagnetic molecules in helium nanodroplets would provide a controllable model  system to study angular momentum dynamics between the electron spin, electron orbital, and lattice degrees of freedom in solids. Such dynamics lies at the core of the Einstein-de Haas and Barnett effects~\cite{EinsteindeHaas15de,BarnettPR15,Dornes2018}, whose detailed quantum-mechanical description is still missing, one century after their discovery.

The authors acknowledge stimulating discussions with Fabian Grusdt, Johan Mentink and Nicol\`o Defenu at various stages of this work. M.L.~acknowledges support by the Austrian Science Fund (FWF), under project No.~P29902-N27, and by the European Research Council (ERC) Starting Grant No.~801770 (ANGULON). G.B.~acknowledges support from the Austrian Science Fund (FWF), under project No.~M2461-N27. I.C.~acknowledges the support by the European Union's Horizon 2020 research and innovation programme under the Marie Sk\l{}odowska-Curie Grant Agreement No.~665385. R.S.~is supported by the Deutsche Forschungsgemeinschaft (DFG, German Research Foundation) under Germany's Excellence Strategy -- EXC-2111 -- 390814868. H.S.~acknowledges support from the European Research Council-AdG (Project No. 320459, DropletControl).

\clearpage
\pagebreak
\widetext

\clearpage
\setcounter{equation}{0}
\setcounter{figure}{0}
\setcounter{table}{0}
\setcounter{page}{1}
\makeatletter
\renewcommand{\theequation}{S\arabic{equation}}
\renewcommand{\thefigure}{S\arabic{figure}}
\renewcommand{\thetable}{S\arabic{table}}

\begin{center}
\textbf{\large Supplemental Material:\\[5pt] Far-from-equilibrium dynamics of angular momentum\\[5pt] in a quantum many-particle system}
\end{center}

\section{Experimental setup and methods}

Details of the experimental setup were described in Ref.~\cite{Shepperson17} and, thus, only a brief description of the relevant details are given here. Helium droplets are produced using a continuous helium droplet source with stagnation conditions of \SI{25}{bar} and \SI{14}{K} for \ce{I2} and \ce{OCS} and \SI{16}{K} for \ce{CS2}, giving $\sim$\SI{10}{nm} diameter helium droplets~\cite{ToenniesAngChem04}. The beam of He droplets exit the source chamber through a skimmer with a \SI{1}{mm} diameter opening and enters a pickup cell containing a vapor of either \ce{CS2}, \ce{OCS} or \ce{I2} molecules. The partial pressure of the molecular vapor was kept sufficiently low to ensure the pickup of at most one molecule per droplet. Hereafter, the doped droplets pass through a liquid nitrogen trap that captures the majority of the effusive molecules that are not picked up by the droplets. In order to further reduce the contribution from effusive molecules the doped droplets pass through a second skimmer with a \SI{2}{mm} diameter opening followed by a second liquid nitrogen trap. Finally, the doped droplets enter a velocity map imaging (VMI) spectrometer placed in the middle of the target chamber. Here, the droplet beam is crossed perpendicularly by two collinear, focused, pulsed laser beams, both with a central wavelength of \SI{800}{nm}

The pulses in the first beam are used to induce alignment. These pulses have a duration of \SI{15}{ps} (the measured temporal intensity profile is shown by the shaded blue shape in each panel in Fig. 1 in the main article) and a Gaussian spotsize, $\omega_\text{0}$ = \SI{30}{\micro\metre}. The pulses, termed probe pulses, in the second beam, sent at time $t$ with respect to the center of the alignment pulses, are used to measure the spatial orientation of the molecules. This occurs by Coulomb explosion of the molecules and recording of the direction of the fragment ions recoiling along the internuclear axis of their parent molecule (\ce{IHe^+} for \ce{I_2}, \ce{S^+} for \ce{CS_2} and  \ce{OCS}). These pulses have a duration of \SI{40}{fs}, spotsize, $\omega_\text{0}$ = \SI{22}{\micro\metre}, and a peak intensity $\SI{8e14}{W/cm^2}$.

The VMI spectrometer projects the ions produced by the probe pulse onto a 2-dimensional detector. The angle between the position of an ion hit and the polarization direction of the alignment beam, contained in the detector plane, is denoted $\theta_\text{2D}$. The degree of alignment is characterized by $\langle \cos^2\theta_\text{2D} \rangle$, a standard measure used in many previous works \cite{Sondergaard:2017id}. The 2-dimensional ion images are recorded at a larger number of delays between the alignment and the probe pulse.  Hereby the time-dependent $\langle \cos^2  \theta_\text{2D} \rangle$ curves, displayed in Fig. 1 in the main article, are obtained.

\section{Theoretical framework: out-of-equilibrium dynamics of angulons in the strong coupling regime}

Our theoretical approach is based on the angulon quasiparticle \cite{Lemeshko2015}, a quantum rotor interacting with many-particle bosonic bath. Originally derived to describe a molecule immersed in a dilute BEC, the theory has been extended to describe phenomenologically molecules trapped inside much denser and strongly-interacting solvents, such as superfluid $^4$He nanodroplets. In particular, angulon theory showed a good agreement in describing experimental observations in helium droplets, namely, renormalization of rotational constants \cite{LemeshkoDroplets16}, impulsive molecular alignment \cite{Shepperson:2017gb,Shepperson17} and selective broadening of spectral lines \cite{Cherepanov17}. Heavy molecules (with rotational constant $B \lesssim 1$~cm$^{-1}$) such as I$_2$, CS$_2$, and OCS considered in the present paper are known to strongly interact with surrounding helium \cite{LemeshkoDroplets16}. This is mostly due to the fact that the kinetic energy of molecular rotation is comparable to the potential energy of the molecule-helium interaction, as opposed to light molecules. Accordingly, we make use of the strong coupling angulon theory developed in Ref. \cite{SchmidtLem16}, accounting for an infinite number of helium excitations.

\subsection{The angulon Hamiltonian}

We start from the Hamiltonian defined in the molecular (body-fixed) coordinate system, co-rotating with the molecule, describing a linear molecule interacting with a bosonic bath
\begin{equation}
\label{eq:Hamiltonian}
\hat{H}=B(\boldsymbol{\mathrm{\hat{L}}}-\boldsymbol{\mathrm{\hat{\Lambda}}})^2 +\sum_{k \lambda \mu} \omega_k \hat{b}^{\dagger}_{k \lambda \mu}\hat{b}_{k \lambda \mu}+\sum_{k \lambda} V_{\lambda}(k)\big(\hat{b}^{\dagger}_{k \lambda 0} +\hat{b}_{k \lambda 0}\big) \; .
\end{equation}
Here we used the notation $\sum_k=\int dk$, set $\hbar \equiv 1$, $B=\hbar^2/(2I)$ is the gas phase rotational constant of the molecule, with $I$ is the molecular moment of inertia, and $\boldsymbol{\mathrm{\hat{L}}}$ the total angular-momentum operator acting in the frame co-rotating with the molecule. Note that the  $\boldsymbol{\mathrm{\hat{L}}}$ operator acts on symmetric top states, since the linear-rotor molecule molecule is turned into an effective symmetric top by dressing of the boson field \cite{SchmidtLem16}. Moreover
\begin{equation}
\boldsymbol{\mathrm{\hat{\Lambda}}}=\sum_{k \lambda \mu \nu} \hat{b}^{\dagger}_{k \lambda \mu} \boldsymbol{\mathrm{\sigma}}^{\lambda}_{\mu \nu}\hat{b}_{k \lambda \nu}
\end{equation}
is the angular-momentum operator for the bosonic helium bath, whose excitations are described by the creation (annihilation) operators, $\hat{b}^{\dagger}_{k \lambda \mu}$ ($\hat{b}_{k \lambda \mu}$), respectively. Furthermore, $\boldsymbol{\mathrm{\sigma}}^{\lambda}=\{\sigma^{\lambda}_{-1},\sigma^{\lambda}_{0},\sigma^{\lambda}_{+1}\}$ denotes the vector of the angular momentum matrices fulfilling the $SO(3)$ algebra in the representation of angular momentum $\lambda$. For convenience, the creation and annihilation operators are cast in the angular momentum basis~\cite{sLemSchmidtChapter}, with $k$ the magnitude of linear momentum, $\lambda$ angular momentum of the boson, and $\mu$ the angular momentum projection onto the laboratory $z$-axis. Finally, $\omega_k$ gives the empirical dispersion relation of superfluid helium~\cite{Donnelly1998}. The details of the molecule-helium interaction are encoded in the potential $V_{\lambda}(k)$
\begin{equation}
\label{eq:expansion}
V_{\lambda}(k)=\sqrt{\frac{n_0 k^4}{\pi m \omega_k}} \int dr r^2 f_{\lambda}(r)j_{\lambda}(kr)
\end{equation}
where $n_0$ is the particle density of helium atoms, $m$ is the mass of a $^4$He atom, and $j_{\lambda}(kr)$ are the spherical Bessel functions. The form factor $f_{\lambda}(r)$ determines the components of the expansion of the molecule-He potential energy surface (PES) into the spherical harmonics. Such a treatment is fully justified only for a low value of helium density. However, we previously demonstrated~\cite{LemeshkoDroplets16,Shepperson:2017gb,Cherepanov17} that fine details of the molecule-helium potential are irrelevant, and the problem can be approached from a phenomenological perspective by scaling the coupling constants, $V_{\lambda}(k)$, according to the particle density in helium. For the sake of simplicity we choose Gaussian form-factors
\begin{equation}
f_{\lambda}=u_{\lambda} (2\pi)^{-3/2}\exp^{-\frac{r^2}{2r^2_{\lambda}}}
\end{equation}
as model potentials. Here $r_{\lambda}$, the interaction range, is set to the distance of the global minimum in the molecular PES, whereas $u_{\lambda}$, the interaction strength, is fixed phenomenologically to reproduce known properties of the molecule-helium interaction,  More details on the model parameters will be given in a subsequent Section. The Hamiltonian (\ref{eq:Hamiltonian}) can be diagonalized in the limit of the slowly rotating molecule $B \to 0$ by means of a coherent-state transformation:
\begin{equation}
\label{eq:trHamiltonian}
\hat{\mathscr{H}}=\hat{U}^{-1} \hat{H} \hat{U}
\end{equation}
where
\begin{equation}
\hat{U}=\exp[\sum_{k\lambda}\frac{V_{\lambda}(k)}{\omega_k+B\lambda(\lambda+1)}(\hat{b}^{\dagger}_{k\lambda0}-\hat{b}_{k\lambda0})] \;.
\end{equation}
After the transformation, the bosonic vacuum $\vert 0\rangle_\text{bos}$ becomes the ground state of the Hamiltonian (\ref{eq:trHamiltonian}). It is also worth noting that for a given total angular momentum state $\vert LM \rangle$, the ground state of the Hamiltonian (\ref{eq:Hamiltonian}), $\ket{\psi_{LM}}=\hat{U}\vert 0 \rangle_\text{bos} \vert LM0 \rangle_\text{mol}$, involves an infinite number of bath excitations and describes the macroscopic deformation of bosons caused by molecular rotation. In a simple picture, one can regard this state as a shell of bosons co-rotating along with the molecule. In the case of a helium nanodroplet playing a role of the many-body bath, such a deformation is known as a nonsuperfluid helium solvation shell \cite{GrebenevOCS}. In the absence of external fields, the system wavefunction can be described by the following time-dependent variational Ansatz
\begin{equation}
\label{eq:ansatz}
\ket{\psi_{LM, i} (t)} =  \hat{U} \biggl ( g_{LM}(t) \ket{LM0}_\text{mol} + \sum_{k \lambda n} \alpha^{LM}_{k \lambda n}(t) \ket{LMn}_\text{mol} \hat{b}^{\dagger}_{k \lambda n} \biggr)  \ket{i}_\text{bos} \; .
\end{equation}
where $\ket{i}$ represents the many-body states of the bath, $L$ and $M$ are constants of motion corresponding to the total angular momentum of the system and its projection onto the laboratory $z$-axis, $n$ defines the projection of the total angular momentum onto the molecular axis, $g_{LM}(t)$ and $\alpha_{k \lambda n}(t)$ are time-dependent variational coefficients. A detailed derivation of the Hamiltonian and its properties can be found in Refs.~\cite{SchmidtLem16,sLemSchmidtChapter}.

\subsection{The Lagrangian}

The Lagrangian of the system in the absence of the laser is
\begin{equation}
\label{eq:Lagrangian}
\mathscr{L}= \frac{1}{Z_\text{bos}} \sum_i e^{-\beta E_i} \langle \psi_{LM,i}(t) \vert i \partial_{t}- \hat{H} \vert \psi_{LM,i}(t) \rangle
\end{equation}
the index $i$ running over the energy eigenstates of the bosonic Fock space, the $\ket{i}$ state having $E_i$ energy, $Z_\text{bos} \equiv \sum_i e^{-\beta E_i}$ being the partition function for the bosonic bath. Here and below, the notation
\begin{equation}
\expval{f(\hat{b}^\dagger,\hat{b})}_\text{T} \equiv \frac{1}{Z_\text{bos}} \sum_i e^{-\beta E_i} \matrixel{i}{f(\hat{b}^\dagger,\hat{b})}{i}
\end{equation}
indicates a finite-temperature bosonic expectation value. For example, one sees immediately that
\begin{equation}
\expval{\hat{b}_{k \lambda n} \hat{b}^\dagger_{k' \lambda' n'}}_\text{T} = \delta(k-k') \delta_{\lambda \lambda'} \delta_{n n'} (1 + f_\text{BE} (\omega_k)) \hspace{32pt} f_\text{BE} (x) = \frac{1}{e^{\beta x} - 1}
\end{equation}
having introduced the Bose-Einstein distribution $f_\text{BE}$, $\beta \equiv (k_B T)^{-1}$. Generalizations to an arbitrary number of bosonic operators are readily found by means of Wick's theorem. Substituting the Hamiltonian (\ref{eq:Hamiltonian}) into Eq. (\ref{eq:Lagrangian}) we obtain
\begin{equation}
\label{eq:Lagrangian-expl-term}
\mathscr{L}= ig^*_{LM}(t)\dot{g}_{LM}(t)+i\sum_{k \lambda n} \alpha^{*LM}_{k \lambda n}(t) \dot{\alpha}^{LM}_{k \lambda n}(t)(1+f_{\text{BE}}(\omega_k))+\mathscr{L}_A+\mathscr{L}_B+\mathscr{L}_C
\end{equation}
where
\begin{equation}
\mathscr{L}_A =- BL(L+1) \vert g_{LM}(t)\vert^2 \; ,
\end{equation}
\begin{align}
\begin{split}
\mathscr{L}_B &=-2B\sum_{k \lambda n} g^*_{LM}(t) \alpha^{LM}_{k \lambda n}(t)\frac{V_{\lambda}(k)}{W_{\lambda}(k)}\sqrt{\lambda(\lambda+1)L(L+1)}\delta_{n \pm1}+ \\
& + 2B\sum_{k\lambda}\lambda(\lambda+1)\alpha^{LM}_{k\lambda0}(t) g^*_{LM}(t)\frac{V_{\lambda}(k)}{W_{\lambda}(k)}(1+f_{\text{BE}}(\omega_k))f_{\text{BE}}(\omega_k) \; , \\
\end{split}
\end{align}
and
\begin{align}
\begin{split}
\mathscr{L}_C&=-\sum_{k \lambda n}BL(L+1) \vert \alpha^{LM}_{k \lambda n}(t) \vert^2(1+f_{\text{BE}}(\omega_k))-\sum_{k \lambda n}\bigg[\omega_k +B\lambda(\lambda+1)\bigg] \vert \alpha^{LM}_{k \lambda n}(t) \vert^2(1+f_{\text{BE}}(\omega_k))^2+\\
& - 2B \sum_{k k' \lambda \lambda' n } \alpha^{*LM}_{k \lambda n}(t)\alpha^{LM}_{k' \lambda' n}(t)\frac{V_{\lambda}(k)}{W_{\lambda}(k)}\frac{V_{\lambda'}(k')}{W_{ \lambda'}(k')} \sqrt{\lambda(\lambda+1)} \sqrt{\lambda'(\lambda'+1)} \delta_{n \pm1}(1+f_{\text{BE}}(\omega_k))(1+f_{\text{BE}}(\omega_{k'}))+\\
&+2B\sum_{k \lambda n \nu} \alpha^{LM}_{k \lambda n}(t) \alpha^{*LM}_{k \lambda \nu}(t)\boldsymbol{\eta}^{L}_{\nu n}\boldsymbol{{\sigma}}^{\lambda}_{\nu n}(1+f_{\text{BE}}(\omega_k))^2 \; .
\end{split}
\end{align}
Note that $W_{\lambda}(k)=\omega_k +B\lambda(\lambda+1)$, and the angular momentum coupling term is given by
\begin{multline}
\label{coupling}
\boldsymbol{\eta}^{L}_{n \nu}\boldsymbol{{\sigma}}^{\lambda}_{n \nu}=n^2 \delta_{n \nu} +\frac{1}{2}\sqrt{\lambda(\lambda+1)- \nu(\nu+1)}\sqrt{L(L+1)- \nu(\nu+1)} \delta_{n \nu+1}+ \\
+ \frac{1}{2}\sqrt{\lambda(\lambda+1)- \nu(\nu-1)}\sqrt{L(L+1)- \nu(\nu-1)} \delta_{n \nu-1}
\end{multline}

\subsection{The equations of motion}

The equations of motion are given by
\begin{equation}
\label{eq:equationsofmotion}
\frac{d}{dt}\frac{\partial \mathscr{L}}{\partial \dot{x_i}}-\frac{\partial \mathscr{L}}{\partial x_i}=0
\end{equation}
where $x_i=\{g_{LM}, \alpha^{LM}_{k \lambda n}\}$. By substituting the Lagrangian (\ref{eq:Lagrangian-expl-term}) into Eq. (\ref{eq:equationsofmotion})  we obtain the system of integro-differential equations for $g_{LM}(t)$ and $\alpha^{LM}_{k \lambda n}(t)$

\begin{subnumcases}{\label{eq:eom}}
\begin{multlined}
 i\dot{g}_{LM}(t)=\bigg[BL(L+1)-D[L(L+1)]^2 \bigg]g_{LM}(t)+ \\
 +2B\sum_{k \lambda n}\frac{V_{\lambda}(k)}{W_{\lambda}(k)}\sqrt{\lambda(\lambda+1)L(L+1)}\delta_{n \pm1} \alpha^{LM}_{k \lambda n}(t)(1+f_{\text{BE}}(\omega_k))+ \\
                  +2B\sum_{k\lambda}\lambda(\lambda+1)\alpha^{LM}_{k\lambda0}(t) \frac{V_{\lambda}(k)}{W_{\lambda}(k)}(1+f_{\text{BE}}(\omega_k))f_{\text{BE}}(\omega_k)
\end{multlined}
 \label{eq:eom1} \\
\begin{multlined} i\dot{\alpha}^{LM}_{k \lambda n}(t) =\bigg[BL(L+1)-D[L(L+1)]^2+(B \lambda(\lambda+1)+\omega_k)(1+f_{\text{BE}}(\omega_k)) \bigg] \alpha^{LM}_{k \lambda n}(t)+\\ + B\frac{V_{\lambda}(k)}{W_{\lambda}(k)}\sqrt{\lambda(\lambda+1)} \sum_{k' \lambda'} \frac{V_{\lambda'}(k')}{W_{\lambda'}(k')} \sqrt{\lambda'(\lambda'+1)} \delta_{n \pm1} \alpha^{LM}_{k' \lambda' n}(t)(1+f_{\text{BE}}(\omega_{k'}))+ \\ + B\frac{V_{\lambda}(k)}{W_{\lambda}(k)}\sqrt{\lambda(\lambda+1)L(L+1)}\delta_{n \pm1}g_{LM}(t)+2B\lambda(\lambda+1) g_{LM}(t)\frac{V_{\lambda}(k)}{W_{\lambda}(k)}\delta_{n0}f_{\text{BE}}(\omega_k) +\\ - 2B \sum_{\nu} \boldsymbol{\eta}^{L}_{n \nu}\boldsymbol{{\sigma}}^{\lambda}_{n \nu}\alpha^{LM}_{k \lambda \nu}(t)(1+f_{\text{BE}}(\omega_k)) \end{multlined} \label{eq:eom2}
\end{subnumcases}
Here we introduced a phemenological term $-D[L(L+1)]^2$, with $D$ the centrifugal distortion constant, accounting for non-rigidity of the system being in the highly-excited total angular momentum states, acting up to a cutoff $L_\text{max}$. For molecules trapped in helium droplets $D$ is four orders of magnitude larger than for the gas phase \cite{LehmannJCP01,sHarms:1997iv,sHartmann:1995cr}. The centrifugal correction to the spectrum becomes non-negligible for $L \gtrsim 5$.

\section{Modelling the interaction of a molecule with a laser pulse}

\subsection{`Laser part' of the Lagrangian}

Since the laser will create a superposition of states with different $L$, an appropriate wavefunction for the system described by the full Hamiltonian given in the main text, including the laser-molecule interaction, is
\begin{equation}
\ket{\Psi_i (t)} = \sum_{LM} \ket{\psi_{LM, i} (t)} \; .
\label{eq:bigpsi}
\end{equation}
The interaction of the molecule with a linearly polarized far-off-resonant laser is modelled, in the laboratory frame of reference, by the following term~\cite{sLemKreDoyKais13}
\begin{equation}
\label{eq:laser}
\hat{H}_\text{laser} = - \frac{1}{4} \Delta \alpha E^2 (t) \cos^2 \hat{\theta} \; .
\end{equation}
One can use the unitary transformation $\hat{S}$ introduced in Ref. \cite{SchmidtLem16} to express it in the molecular frame of reference where the angulon Hamiltonian (\ref{eq:Hamiltonian}) and the variational Ansatz (\ref{eq:bigpsi}) are defined, obtaining $\hat{\mathscr{H}}_\text{laser} = \hat{S}^{-1} \hat{H}_\text{laser} \hat{S}$. Then, the molecular-laser interaction will enter the Lagrangian of the system as the additional term
\begin{equation}
\mathscr{L}_\text{laser} = - \frac{1}{Z_\text{bos}} \sum_i e^{-\beta E_i} \matrixel{\Psi_i (t)}{\hat{\mathscr{H}}_\text{laser}}{\Psi_i (t)},
\end{equation}
which yields the `laser Lagrangian'
\begin{equation}
\mathscr{L}_\text{laser}  = \frac{1}{4} \Delta \alpha E^2 \int \mathrm{d} \Omega \cos^2 \theta \frac{1}{Z_\text{bos}} \sum_i e^{-\beta E_i} \braket{\Psi_i (t)}{\Omega} \braket{\Omega}{\Psi_i (t)} \; .
\label{eq:lagr1}
\end{equation}
The selection rules for the Clebsch-Gordan coefficients imply that $M=M'$, as expected since the electric field is conserving the $z$-component of angular momentum. This means that the time evolution does not mix states with different values of $M$. After a straightforward calculation one gets that the time-evolution on a manifold with defined $M$ is described by the following Lagrangian
\begin{multline}
\mathscr{L}_\text{laser} = \frac{1}{6} \Delta \alpha E^2 ( \sum_{L L' } g^*_{LM} g_{L'M} \sqrt{\frac{2 L + 1}{2L'+1}} C^{L'M}_{20LM} C^{L'0}_{20L0} + \\ + \sum_{L L' } \sum_{k \lambda n} \alpha^{LM \ *}_{k \lambda n} (t) \alpha^{L'M}_{k \lambda n} (t) (1 + f_\text{BE} (\omega_k)) \sqrt{\frac{2 L + 1}{2L'+1}} C^{L'M}_{20LM} C^{L'n}_{20Ln} + \\
+ \frac{1}{2} \sum_{L} g^*_{LM} g_{LM} + \frac{1}{2} \sum_{L} \sum_{k \lambda n} \alpha^{LM \ *}_{k \lambda n} (t) \alpha^{LM}_{k \lambda n} (t) (1 + f_\text{BE} (\omega_k)))
\end{multline}
via the corresponding Euler-Lagrange equations of motion.

\subsection{`Laser part' of the equations of motion}

In order to simplify the notation let us define
\begin{equation}
C \equiv \frac{1}{6} \Delta \alpha E^2 \hspace{32pt} Q_{L L' M N} \equiv \sqrt{\frac{2 L + 1}{2L'+1}} C^{L'M}_{20LM} C^{L'N}_{20LN}
\end{equation}
so that
\begin{multline}
\mathscr{L}_\text{laser} = C ( \sum_{L L'} g^*_{LM} g_{L'M} Q_{L L' M 0} + \sum_{L L'} \sum_{k \lambda n} \alpha^{LM \ *}_{k \lambda n} (t) \alpha^{L'M}_{k \lambda n} (t) (1 + f_\text{BE} (\omega_k)) Q_{L L' M n} + \\
+ \frac{1}{2} \sum_{L} g^*_{LM} g_{LM} + \frac{1}{2} \sum_{L} \sum_{k \lambda n} \alpha^{LM \ *}_{k \lambda n} (t) \alpha^{LM}_{k \lambda n} (t) (1 + f_\text{BE} (\omega_k)) )
\end{multline}
Then
\begin{equation}
\frac{\partial \mathscr{L}_\text{laser}}{\partial g_{LM}^*} = C \sum_{L'} Q_{L L' M 0} \ g_{L'M} + \frac{C}{2} g_{LM}
\label{eq:elg}
\end{equation}
and
\begin{equation}
\frac{\partial \mathscr{L}_\text{laser}}{\partial \alpha^{LM*}_{k \lambda n}} = C \sum_{L' \geq 2} Q_{L L' M n} \ \alpha^{L'M}_{k \lambda n} (1 + f_\text{BE} (\omega_k)) + \frac{C}{2} \alpha^{LM}_{k \lambda n} (1 + f_\text{BE} (\omega_k))
\label{eq:elalpha}
\end{equation}
to be added to the equations of motion of Eq. (\ref{eq:eom}). It is important noting that, due to the selection rules imposed by $Q_{L L' M n}$, the summation on $L'$ of Eq. (\ref{eq:elg}) and Eq. (\ref{eq:elalpha}) extends only on $L'=L-2,\ldots,L+2$. Also in Eq. (\ref{eq:elalpha}) one needs to introduce the condition $L' \geq 2$ since the phonon amplitudes $\alpha^{LM}_{k \lambda n}$ vanish for $L \leq 1$ \cite{SchmidtLem16}.

\section{Calculation of the projected cosine, $\cos^2 \theta_{2D}$}

The experimentally-measured molecular alignment is defined in terms of the operators describing the molecular angles $\hat{\theta}$ and $\hat{\phi}$ as
\begin{equation}
\cos^2 \hat{\theta}_{2D} = \frac{\cos^2 \hat{\theta}}{\cos^2 \hat{\theta} + \sin^2 \hat{\theta} \sin^2 \hat{\phi}} \; .
\label{eq:expval0}
\end{equation}
In order to calculate the matrix elements of this operator in the angular momentum basis $\matrixel{j'm'}{\cos^2 \hat{\theta}_{2D}}{jm}$ we expand it as follows
\begin{equation}
\cos^2 \hat{\theta}_{2D} = \sum_{\lambda \mu} f_{\lambda \mu} Y_{\lambda \mu} (\hat{\theta},\hat{\phi})
\label{eq:expval1}
\end{equation}
and clearly the inverse expansion, giving the coefficients $f_{\lambda \mu}$ in terms of $\cos^2 \theta_{2D}$, is
\begin{equation}
f_{\lambda \mu} = \int \mathrm{d} \Omega \ Y_{\lambda \mu} (\Omega) \cos^2 \theta_{2D} \; .
\label{eq:fdef}
\end{equation}
Here the idea is that, rather than dealing with the cumbersome expectation value in Eq. (\ref{eq:expval0}), one can calculate an infinite series of expectation values of the type $\matrixel{j'm'}{Y_{\lambda \mu}(\hat{\theta},\hat{\phi})}{j'm'}$ -- inserting the expansion in Eq. (\ref{eq:expval1}) -- which will be effectively limited by an angular momentum cutoff. The calculation of the expectation value $\matrixel{j'm'}{\cos^2 \hat{\theta}_{2D}}{jm}$ is then simplified to
\begin{multline}
\matrixel{j'm'}{\cos^2 \hat{\theta}_{2D}}{jm} =  \sum_{\lambda \mu} f_{\lambda \mu} \matrixel{j'm'}{Y_{\lambda \mu} (\hat{\theta},\hat{\phi})}{jm} = \\ = \sum_{\lambda \mu} f_{\lambda \mu} (-1)^{m'} \sqrt{\frac{(2j+1) (2j'+1) (2\lambda+1)}{4 \pi}} \begin{pmatrix} j & j' & \lambda \\ 0 & 0 & 0 \end{pmatrix} \begin{pmatrix} j & j' & \lambda \\ m & -m' & \mu \end{pmatrix} \; .
\label{eq:cos2d0}
\end{multline}
We still have to calculate $f_{\lambda \mu}$ through Eq. (\ref{eq:expval0}). Since, as already noted, the laser does not break the symmetry for rotations around the $z$-axis, one will always have to deal with the $m=m'$ case, which implies $\mu=0$ in Eq. (\ref{eq:cos2d0}). Then the coefficients $f_{\lambda 0}$ are given by
\begin{equation}
f_{\lambda 0} = \sqrt{\frac{2 \lambda + 1}{4 \pi}} \int_{-1}^{+1} \mathrm{d} x \int_0^{2 \pi} \mathrm{d} \phi \ P_\lambda (x) \frac{x^2}{x^2 + (1-x^2) \sin^2 \phi} \; .
\end{equation}
By noting that $\int_0^{2 \pi} \frac{\mathrm{d} \phi}{x^2 + (1-x^2)\sin^2 \phi} = \frac{2 \pi}{|x|}$ one gets
\begin{equation}
f_{\lambda 0} = 4 \pi \sqrt{\frac{2 \lambda + 1}{4 \pi}} \int_0^1 \mathrm{d} x P_\lambda (x) P_1 (x) \hspace{32pt} \text{if } \lambda \text{ is even}
\label{eq:flambda0}
\end{equation}
whereas for odd $\lambda$ one sees that $f_{\lambda0}=0$ due to symmetry arguments. The integral in Eq. (\ref{eq:flambda0}) can be evaluated in closed a form \cite{Byerly:1893} -- note that it is \textit{not} the orthogonality relation for Legendre polynomials -- and it finally leads to the following closed form for $f_{\lambda 0}$,
\begin{equation}
f_{\lambda 0} = \begin{cases}
4 \pi \sqrt{\frac{2 \lambda + 1}{4 \pi}} \frac{(-1)^{\frac{\lambda}{2} + 1} \lambda!}{2^\lambda (\lambda - 1) (\lambda + 2) (\frac{\lambda}{2} !)^2} & \lambda \mbox{ even} \\
0 & \lambda \mbox{ odd}
\end{cases}
\end{equation}

\section{Ensemble and focal averaging}

\newcommand{\myexpval}[1]{\langle #1 \rangle}

After calculating the time-evolution of the wavefunction $\ket{\Psi_i (t)}$ by solving the Euler-Lagrange equations of motion, one may want to calculate expectation values for different observables. For concreteness' sake let us consider a certain solution to the equations of motion $\ket{\Psi_i (t)}_{L_0,M_0}$ obtained starting from an initial purely-molecular wavefunction $\ket{\Psi_i (t=0)}=\ket{LM0}$, that will be dynamically dressed; let us also consider a generic operator $\hat{O}$. Taking the expectation value of $\hat{O}$ following Ref. \cite{sDeAngelis:1991dj}, one accounts for the finite temperature of the bosonic environment as
\begin{equation}
\myexpval{\hat{O}}_{L_0, M_0} (t) = \frac{1}{Z_\text{bos}} \sum_i e^{-\beta E_i} \matrixel{\Psi_i (t)}{\hat{O}}{\Psi (t)}_{L_0,M_0} \; .
\end{equation}
However, the energy scale corresponding to the rotational constant for all the three molecules we consider is considerably smaller than the temperature of the bath $T_\text{bath}=0.38 \text{K}$, so that an expectation value should be calculated as an average over a statistical ensemble. The occupation probabilities for each state in the ensemble are given by the normalised Boltzmann factors \cite{Shepperson:2017gb}
\begin{equation}
P_{LM} = \frac{e^{- \beta B^* L(L+1)}}{Z}
\label{eq:plm}
\end{equation}
with $Z=\sum_{LM} P_{LM}$, $B^*$ being the effective rotational constant as derived in Ref.~\cite{LemeshkoDroplets16}, $\beta=(k_B T)^{-1}$. In addition to this, one also has to take into account the fact, generally, the even and odd states will have different relative abundances. To this extent, the occupation probabilities of Eq. (\ref{eq:plm}) need to be calculated separately for the even and odd components, as
\begin{equation}
P^\text{even}_{LM} = \frac{e^{- \beta B^* L(L+1)}}{Z^\text{even}} \hspace{32pt} Z^\text{even}=\sum_{L \text{ even}, M} P^\text{even}_{LM}
\end{equation}
and similarly for $P^\text{odd}_{LM}$ and $Z^\text{odd}$. Practically, both partition functions can be calculated up to some angular momentum cutoff $L_\text{max}$, for the molecules we consider in the present paper a cutoff $L_\text{max}=8$ which suffices to always include more than $99\%$ of the ensemble. Going back to the expectation value of the operator $\hat{O}$, we define the averages over the even and odd components as
\begin{equation}
\myexpval{\hat{O}}_\text{even} (t) = \sum_{\substack{L_0,M_0\\L_0 \text{ is even}}} P^\text{even}_{L_0,M_0} \myexpval{\hat{O}}_{L_0, M_0} (t)
\end{equation}
and
\begin{equation}
\myexpval{\hat{O}}_\text{odd} (t) = \sum_{\substack{L_0,M_0\\L_0 \text{ is odd}}} P^\text{odd}_{L_0,M_0} \myexpval{\hat{O}}_{L_0, M_0} (t) \; .
\end{equation}
Introducing the abundances of the even and odd states, let us call them $A_\text{even}$ and $A_\text{odd}$, respectively, one gets:
\begin{equation}
\myexpval{\hat{O}} (t) = \frac{A_\text{even} \myexpval{\hat{O}}_\text{even} (t) + A_\text{odd} \myexpval{\hat{O}}_\text{odd} (t)}{A_\text{even} + A_\text{odd}}
\end{equation}
which corresponds to the expectation value of the operator $\hat{O}$ taking into account finite-temperature effects for the bath and for the molecule. Finally, we performed a weighted average of five different peak intensities defined by the measured spot sizes of the alignment and the probe pulse laser beams.

\section{Transition probability under a Gaussian time-dependent perturbation}

Here we derive the transition probability $W_{L,L'}$ given in the main text. Let us consider a time-dependent perturbation whose time-dependence is Gaussian
\begin{equation}
\hat{V} (t) = \hat{V} \frac{1}{\sqrt{2 \pi \sigma^2}} \mathrm{exp} (-t^2/(2 \sigma^2)) \; .
\end{equation}
The probability of transition from an initial stationary state $\ket{i}$ to a final stationary state $\ket{f}$ under the action of $\hat{V} (t)$ is given by \cite{Landau:1981}
\begin{equation}
W_{fi} = \frac{1}{\hbar^2} \left| \int_{-\infty}^{+\infty} V_{fi}(t) e^{\mathrm{i} \omega_\text{fi} t} \mathrm{d} t \right|^2
\end{equation}
where $\omega_\text{fi} \equiv (E_f - E_i)/\hbar$ and $V_{fi} (t)$ is a matrix element of $\hat{V} (t)$. After carrying out the integral over $\mathrm{d} t$ one has
\begin{equation}
W_{fi} = \frac{|V_{fi}|^2}{\hbar^2} \exp(-\sigma^2 \omega_\text{fi}^2)
\end{equation}
and using $L'$, $L$ as initial (final) states, respectively, one has the expression for $W_{L,L'}$ given in the main text.

\section{Parameters of the model}

For most molecules $f_{\lambda = 2}(r)$ is the dominant anisotropic term in the molecule-helium PES expansion (\ref{eq:expansion})~\cite{Prosmiti:2004cb,Garcia-Gutierrez2009,Farrokhpour2013,Zang2014,Gianturco2000,Ting2014}. Therefore, in order to simplify the final equations of motion (\ref{eq:eom}), one can keep only the many-body excitations with $\lambda=2$. Such an approach has been successfully employed for predicting renormalization of rotational constants leading to a good quantative agreement with experimental data \cite{LemeshkoDroplets16}. All parameters used for simulations of molecular dynamics  are listed in Table \ref{Tab:Tcr}. The constant accounting for centrifugal correction, $D$, is fixed such as to reproduce experimentally the measured value of $D^*$ for OCS and calculated by the following empirical relationship for I$_2$ and CS$_2$ \cite{Choi2006}.
\begin{equation}
D^*=0.0310(38) B^{*1.818(39)}
\end{equation}
the formula being valid for $B^*$ and $D^*$ measured in cm$^{-1}$. We set $r_2$, the parameter characterizing the range of the molecule-helium interaction, using the position of the global minimum of the molecule-He PES as a reference. The strength of  interaction, $u_2$, was chosen as to reproduce the renormalization of rotational constants in helium droplet, $B^*/B$. The particle density of helium atoms in the center of the droplet and their temperature are 0.022~\AA $^{-3}$ and $0.38$~K, respectively \cite{ToenniesAngChem04}. The mass of a $^4$He atom is $m=4.03$~amu.

\begin{table}[h]
\def\arraystretch{1.5}
\begin{tabular}{|c|c|c|c|c|c|c|c|}
\hline
    & $B$ (GHz)& $B^*/B$ & $D^*$ (GHz) & $u_2$ (THz) & $r_2$ (\AA) & $\Delta \alpha$ (\AA$^3$)& Spin abundance \\[-6pt]
    & & & & & & & (even:odd) \\
\hline
I$_2$  & 1.12 \cite{Barrow1973}  &0.6 \cite{Shepperson:2017gb} & $9.00 \cdot 10^{-4}$ &3.14   &  4.8 \cite{Prosmiti:2004cb,Garcia-Gutierrez2009}  &       6.1  \cite{Maroulis1997}    &       15:21          \\
\hline
CS$_2$ & 3.27 \cite{Walker:1971jf} &0.3 \cite{zillich_private_2018}& $2.67 \cdot 10^{-3} $ & 1.05  &  3.4 \cite{Farrokhpour2013,Zang2014,Ting2014} &      10.3 \cite{Miller1990}       &      1:0           \\
\hline
OCS & 6.08 \cite{Hunt1985} &0.36 \cite{GrebenevOCS}& $1.14 \cdot 10^{-2}$ \cite{GrebenevOCS} &  0.50  &  3.8 \cite{Gianturco2000} &       4.7 \cite{dimitrovski_ionization_2011}     &        1:1         \\
\hline
\end{tabular}
\caption{Parameters used in the simulations. \label{Tab:Tcr}}
\end{table}

\section{Time-dependent alignment for OCS molecules}

\begin{figure}[h]
\includegraphics[width=0.4\linewidth]{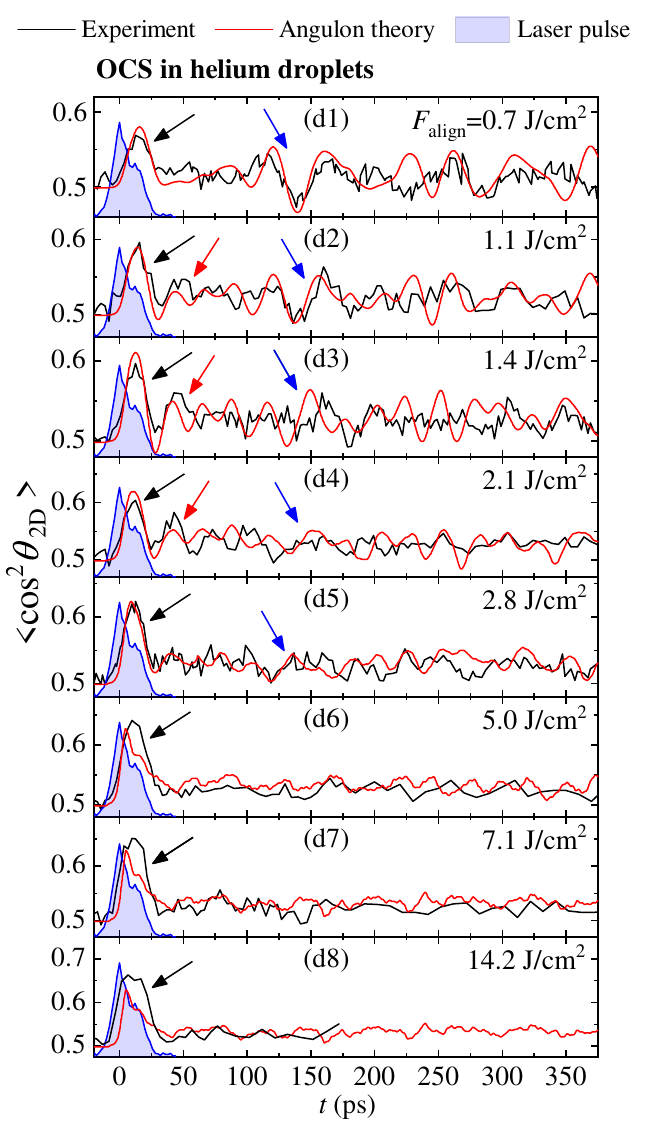}
\caption{\label{fig:OCS} Time evolution of the  degree of alignment, $\langle \cos^2  \theta_\text{2D} \rangle$, for OCS induced by a \SI{15}{ps} laser pulse (whose  intensity profile is shown by the shaded blue shape)  with fluence $F_\text{align}$.   Experiment (black solid lines) is compared to   finite-temperature angulon theory (red solid lines). }
\end{figure}

Here we present the experimental and theoretical results for the time-dependent alignment of OCS, analogous to that of Fig.~1 of the main text. Figure~\ref{fig:OCS} shows $\langle \cos^2 \theta_\text{2D} \rangle$ as a function of time for \ce{OCS} and a series of different fluences of the alignment pulse, $F_\text{align}$. Here, the oscillations are observed at intermediate fluences, most clearly at $F_\text{align} = \SI{2.1}{J/cm^2}$.  Compared to \ce{I_2} and \ce{CS_2} (Fig.~1 of the main text), however, the oscillations  are much less pronounced and disappear around $F_\text{align} = \SI{5.0}{J/cm^2}$. One can see that the results of the angulon quasiparticle theory (red solid line) reproduce all the main features observed in experiment: the peak of prompt alignment (black arrows), weak oscillations (red arrows), as well as the revival structure (blue arrows).

\clearfmfn

\end{document}